\documentclass{elsart}
\usepackage{epsfig,amssymb,amsfonts,amsmath}

\begin{document}
\def\eeq{\end{equation}}
\def\beq{\begin{equation}}
\def\bea{\begin{eqnarray}}
\def\eea{\end{eqnarray}}
\begin{frontmatter}

\title{On the way towards a generalized entropy maximization procedure}

\author{G. Baris Ba\u{g}c\i}
\and
\author{Ugur Tirnakli\corauthref{cor}}
\corauth[cor]{Corresponding author.}
\ead{ugur.tirnakli@ege.edu.tr}
\address{Department of Physics,
Faculty of Science, Ege University, 35100 Izmir, Turkey}

\begin{abstract}
We propose a generalized entropy maximization procedure, which takes
into account the generalized averaging procedures and information
gain definitions underlying the generalized entropies. This novel
generalized procedure is then applied to R\'{e}nyi and Tsallis
entropies. The generalized entropy maximization procedure for
R\'{e}nyi entropies results in the exponential stationary
distribution asymptotically for $q\in \lbrack 0,1]$ in contrast to
the stationary distribution of the inverse power law obtained
through the ordinary entropy maximization procedure. 
Another result of the generalized entropy maximization procedure 
is that one can naturally obtain all the possible stationary distributions 
associated with the Tsallis entropies by employing either ordinary 
or $q$-generalized Fourier transforms in the averaging procedure.
\end{abstract}

\end{frontmatter}

\date{\today}
\maketitle

\section{Introduction}
The inverse power law distributions are ubiquitous in nature,
emerging in such diverse fields as subregion laser cooling [1], the
heartbeat histograms of healthy patients [2], plasmas [3],
conservative motion in 2-D periodic potentials [4], controlled
decoherence [5], rheology of steady-state draining foams [6], DNA
slippage step-length distributions [7], econophysics [8], earthquake
models [9], to name but a few.

On the other hand, it is well-known that the stationary distribution
obtained from the maximization of the Boltzmann-Gibbs (BG) entropy
is of exponential form. In this sense, it cannot always be used in modeling
phenomena exhibiting inverse power law behavior. Therefore, there
has been an increasing interest in the generalized entropies such as
Tsallis [10], R\'{e}nyi [11] and Sharma-Mittal (SM) [12] entropies
whose stationary solutions are of inverse power law form. Although
these different entropy measures yield to inverse power law
stationary distributions, they differ from one another in many
aspects. For example, R\'{e}nyi entropy is additive whereas Tsallis
entropy is not. Both Tsallis and SM entropies are nonadditive, but
the former is one parameter generalization whereas the latter is a
two-parameter generalization of BG entropy. However, one common
structure underlying these generalized entropies is that all of them
are obtained through the joint generalization of the averaging
procedures and the concept of information gain. For example,
R\'{e}nyi entropy preserves the same definition of information gain
as BG measure, but makes use of a different averaging procedure than
the one used for BG entropy, namely, exponential averaging procedure. 
Although this averaging procedure seems to stem from information-theoretic 
approaches, it has wide range of applications in nonequilibrium statistical 
physics. For example, the free energy difference for arbitrary transformations 
is nothing but the exponential average of total work done during the process 
as given by Jarzynski equality [13]. This equality is the cornerstone of 
experimental investigations in several diverse fields [14].

On the other hand, Tsallis measure preserves the same averaging procedure as BG
measure i.e., linear averaging, but generalizes the concept of
information gain by deforming the logarithmic function. The SM
entropy benefits generalizations of both kind in its structure [15].

Although the mathematical structure of these generalized entropies
is well understood, the entropy maximization procedure (EMP) applied
to these generalized measures does not take the aforementioned
structure into account, in general. In fact, the choice of
constraints mostly relies on trial and error or the arguments of
suitability in order to obtain a stationary distribution of inverse
power law. For example, it is emphasized in the literature that
Tsallis and R\'{e}nyi entropies are monotonic functions of one
another and therefore yield the same stationary distribution under
the same constraints. However, there is no reason to apply the same
set of constraints to both, since their mathematical structure is
completely different. At this point, it is worth remark that one can
obtain a stationary distribution of inverse power law form even by
using BG entropy with \textit{suitably} chosen constraints [16].
However, we discard this possibility, since one cannot justify the
form of constraint necessary for this maximization in a reasonable
manner [16, 17].  In short, it has been an open problem how the
constraints for the generalized entropies must be chosen (see Ref.
[18] for the choice of constraints and the nonadditive formalism).

In this paper, we propose a new EMP, which takes into account the
general mathematical structure of the generalized entropy measures
in terms of the underlying definition of the information gain and
generalized averaging procedure. This novel procedure will be
hereafter called generalized entropy maximization procedure (GEMP).

The outline of the paper is as follows. In section II, we review the
mathematical structure of the R\'{e}nyi entropy and generalized
averaging procedures. The application of GEMP to R\'{e}nyi entropy
is presented in Section III. Section IV is the application of GEMP
to Tsallis entropy. Conclusions are presented in Section V.

\section{Generalized averaging procedures and R\'{e}nyi entropy}
It is well-known that BG entropy is the linear average of the
elementary information gain $\log _{b}(1/p_{i})$ associated
with an event of probability $p_{i}$ i.e.,

\begin{equation}
S_{BG}(p)=\left\langle \log _{b}\left( \frac{1}{p_{i}}\right)
\right\rangle _{\text{lin}}
\end{equation}

\noindent where the linear average is defined as

\begin{equation}
\langle x_{i}\rangle _{\text{lin}}\equiv\sum_{i}^{W}p_{i}x_{i}
\end{equation}

\noindent $W$ being the total number of configurations of the
system. It should be noted that from here on we will use the natural
base i.e., $b=e$ and denote $\log _{e}(1/p_{i})$ as
$\ln(1/p_{i})$ without loss of generality. In  order to generalize BG entropy, A.
R\'{e}nyi considered whether other forms of averaging procedures are
possible or not. He then adopted the generalized averaging procedure
developed by Kolmogorov and Nagumo [19, 20]. Kolmogorov and Nagumo
independently showed that the averaging procedure must be extended
to quasi-linear mean defined as

\begin{equation}
S=f^{-1}\left[ \sum_{i}^{W}p_{i}f\left( \ln \frac{1}{p_{i}}\right)
\right]
\end{equation}

\noindent where $f$ is a strictly monotone continuous and invertible
function called Kolmogorov-Nagumo function (K-N function). The
importance of this extension of the averaging procedure is
understood, since it succeeds the generalization by preserving
conformity to Kolmogorov axioms of probability as shown by
Kolmogorov and Nagumo. R\'{e}nyi then showed that \textit{only} two
possible K-N functions exist if one is restricted to additive
measures i.e., for two systems described by two independent
probability distributions A and B, the entropy measure satisfies
$S(A\cup B)=S(A)+S(B\mid A)$, where the conditional probability
$S(B\mid A)$ is defined as $S(B\mid A)=\sum_{i}p_{i}(A)S(B\mid
A=A_{i})$. The first possible K-N function is the linear mean
defined in Eq. (2) and reads $f(x)=x$. The linear mean of the
information gain results in the BG entropy i.e., Eq. (1). The second
possibility is the exponential averaging defined by

\begin{equation}
f(x)=c_{1}e^{(1-q)x}+c_{2}
\end{equation}

\noindent where $q$ is a real parameter, and $c_{1}$,  $c_{2}$ are
two arbitrary constants [11, 15]. The arbitrary constants $c_{1}$
and $c_{2}$ can be chosen, without loss of generality, as
$\frac{1}{1-q}$ and $\frac{1}{q-1}$, respectively. The general
expression for the exponential average of a quantity, using Eqs. (3)
and (4), can then be written as

\begin{equation}
\langle x_{i}\rangle _{\exp }=\frac{1}{1-q}\ln \left(
1+\sum_{i}p_{i}e^{(1-q)x_{i}}-\langle 1\rangle _{\text{lin}}\right).
\end{equation}

\noindent The expression $\langle 1\rangle _{\text{lin}}$ is equal
to one (more on this later) so that

\begin{equation}
\langle x_{i}\rangle _{\exp }=\frac{1}{1-q}\ln \left(
\sum_{i}p_{i}e^{(1-q)x_{i}}\right).
\end{equation}

\noindent Now, it is not difficult to see that the exponential
average of the ordinary information gain results in the R\'{e}nyi
entropy [11, 15] i.e.,

\begin{equation}
S_{R}(p)=\left\langle \ln \left( \frac{1}{p_{i}}\right)
\right\rangle _{\exp}=\frac{1}{1-q}\ln \left(
\sum_{i}^{W}p_{i}^{q}\right)
\end{equation}

\noindent where $\langle \cdot\rangle _{\exp }$ stands for the
exponential averaging procedure defined by Eq. (6). Since the
exponential average becomes the linear average in the
$q\rightarrow1$ limit, the R\'{e}nyi entropy becomes the BG
entropy in the same limit, i.e., as $q\rightarrow1$. In other words,
the only difference between the R\'{e}nyi and BG entropies is due to
the different averaging procedure used although the same definition
of the information gain $\ln(1/p_{i})$ is preserved in both
measures.

Note that, through Eq. (5), the exponential average of $1$ can be written as

\begin{equation}
\langle 1\rangle _{\exp }=\frac{1}{1-q}\ln \left[ 1+e^{(1-q)}\langle
1\rangle _{\text{lin}}-\langle 1\rangle _{\text{lin}}\right].
\end{equation}

\noindent The above equality shows us that once the normalization
through linear averaging procedure is carried out i.e., $\langle
1\rangle _{\text{lin}}=1$, the normalization through exponential
averaging procedure i.e., $\langle 1\rangle _{\exp}=1$ is ensured,
and vice versa.

At this point, it is worth mentioning that if one sets the arbitrary constants 
in Eq.~(4) as $c_1=1$, $c_2=0$ and $1-q=-\beta$, then the Jarzynski 
equality $\Delta F=-\beta \ln\left<\exp(-\beta W)\right>_{\text{lin}}$ [13] 
can simply be written as $\Delta F=\left<W\right>_{\text{exp}}$.

\section{GEMP and R\'{e}nyi entropy}
Since the seminal work of Jaynes, the entropy maximization procedure
played an important role in obtaining the stationary distribution
associated with a particular entropy measure. For example, the
maximization of the BG entropy has been carried out by using the
following functional

\begin{equation}
\Phi _{BG}=\left\langle \ln \frac{1}{p_{i}}\right\rangle _{\text{lin}%
}-\alpha \left\langle 1\right\rangle _{\text{lin}}-\beta
\left\langle \varepsilon _{i}\right\rangle _{\text{lin}}
\end{equation}

\noindent to obtain the concomitant stationary distribution

\begin{equation}
p_{i}=\exp (-S_{BG}+\beta U-\beta \varepsilon _{i}),
\end{equation}

\noindent denoting $\langle \varepsilon _{i}\rangle _{\text{lin}}$
by $U$, as usual. The stationary distribution in Eq. (10) can be cast 
into the form $p_{i}=\frac{e^{-\beta \varepsilon _{i}}}{Z}$, where the 
partition function is given by $Z=e^{S_{BG}-\beta U}$. 
It should be noted that the linear averages in Eq. (9) are carried out 
using ordinary probability distribution (e.g., not escort distribution), 
since ordinary information gain written in
terms of ordinary logarithmic function is used in the definition of
BG entropy. This fact can be better understood by remembering that a
function $f(x)$ is completely and uniquely determined by its moments
when the moments are calculated in terms of the ordinary probability
distribution and with the help of ordinary Fourier transform.
Although the R\'{e}nyi entropy preserves the same definition of
ordinary information gain (and must therefore be maximized using
ordinary probability distribution), the averaging procedure is
exponential rather than linear. However, the maximization of the
R\'{e}nyi entropy too is done exactly in the same way as in BG
entropy in the literature [21-29] i.e.,

\begin{equation}
\Phi _{R}=\left\langle \ln \frac{1}{p_{i}}\right\rangle
_{\text{exp}}-\alpha \left\langle 1\right\rangle _{\text{lin}}-\beta
\left\langle \varepsilon _{i}\right\rangle _{\text{lin}},
\end{equation}

\noindent which is \textit{completely inconsistent}, since the same
averaging procedure is not used throughout the functional to be
maximized. Instead, a consistent maximization would require the use
of the following functional

\begin{equation}
\Phi _{R}=\left\langle \ln \frac{1}{p_{i}}\right\rangle
_{\text{exp}}-\alpha \left\langle 1\right\rangle _{\text{exp}}-\beta \left\langle
\varepsilon _{i}\right\rangle _{\text{exp}}.
\end{equation}

\noindent The maximization of this consistent functional yields

\begin{equation}
\frac{\delta \Phi _{R}}{\delta p_{i}}=\frac{qp_{i}^{q-1}}{%
\sum\limits_{j}p_{j}^{q}}-\alpha \left( 1-e^{q-1}\right) -\beta
\left[ e^{(1-q)(\varepsilon _{i}-U)}-e^{(q-1)U}\right] =0
\end{equation}

\noindent where $U$ is consistently calculated through
$\langle\varepsilon _{i}\rangle _{\exp }$. After a little algebra,
one obtains the stationary distribution

\begin{equation}
p_{i}=\left[ \alpha
\frac{\sum\limits_{j}p_{j}^{q}}{q}(1-e^{q-1})+\beta
\frac{\sum\limits_{j}p_{j}^{q}}{q}\left( e^{(1-q)(\varepsilon
_{i}-U)}-e^{(q-1)U}\right) \right] ^{\frac{1}{q-1}}.
\end{equation}

\noindent In order to eliminate the Lagrange multiplier $\alpha$, we
multiply Eq. (13) by $p_{i}$ and sum over the index $i$ to obtain

\begin{equation}
\alpha =\frac{q+\beta \left[ e^{(q-1)U}-1\right] }{1-e^{q-1}}.
\end{equation}

\noindent By substituting $\alpha$ given by Eq. (15) into Eq. (14),
we obtain the stationary distribution associated with the R\'{e}nyi
entropy measure as

\begin{equation}
p_{i}=\frac{[1-\beta ^{\ast }(1-e^{(1-q)(\varepsilon _{i}-U%
)})]_{+}^{\frac{1}{(q-1)}}}{Z_{q}}
\end{equation}

\noindent with the notation $[a]_{+}=\max \{0,a\}$. The parameter
$\beta ^{\ast }$ is equal to $\frac{\beta }{q}$, and the partition
function $Z_{q}$ is given by
$Z_{q}=(\sum_{j}p_{j}^{q})^{\frac{1}{(1-q)}}$. This stationary
distribution asymptotically decays as an exponential i.e.,
$p_{i}\propto e^{-\varepsilon _{i}}$ for $q\in \lbrack 0,1]$. This
result is in complete agreement with the one obtained from the
method of multinomial coefficients used by Oikonomou [30], and is
different than the one obtained through ordinary EMP, since the
latter obtains a stationary distribution of inverse power law form
[21-29]. We emphasize again that the ordinary maximization of the
R\'{e}nyi entropy is based on the functional in Eq. (11) and
therefore inconsistent.

The analysis of the stationary distribution given by Eq. (16) shows
that it asymptotically attains a constant value, independent of the
microstate energy $\varepsilon _{i}$, for $q>1$. Therefore, the
R\'{e}nyi measure cannot be used as an entropy in a thermostatical
sense for the region $q>1$. This fact has previously been agreed
upon due to the fact that the R\'{e}nyi entropy is neither concave
nor convex in the aforementioned interval. However, a consistent
maximization of the R\'{e}nyi measure simply shows the physical
reason at the level of the stationary distribution (i.e., Eq. (16)),
why this measure cannot be used for $q>1$. The stationary solution
obtained through ordinary EMP of the R\'{e}nyi measure, on the other
hand, does not show such an inconsistency for the interval $q>1$.
This result also verifies the recent findings of Oikonomou and
Tirnakli given in [31], where they mentioned that the inverse power
law distributions obtained from the ordinary EMP for R\'{e}nyi
entropies do not make these entropies extensive. Thus, these
probability distribution functions do not maximize R\'{e}nyi
entropies. Finally, we note that the stationary distribution given
by Eq. (16) becomes the one given by BG entropy i.e., Eq. (10) in
the $q\rightarrow 1$ limit.

\section{GEMP and Tsallis entropy}
We will now work in the continuous domain so that the summation of
the previous section will be replaced by integral. As we have noted
before, generalized entropies is based on the joint generalization
of the concept of information gain and the averaging procedure. The
R\'{e}nyi entropy (which is additive like BG entropy) preserves the
ordinary definition of the information gain, but generalizes the
averaging procedure. Tsallis entropy (which is nonadditive) is
another generalized entropy measure, and is based on the
generalization of the information gain \textit{only}, since it
preserves the linear averaging procedure of the BG entropy. Tsallis
entropy reads

\begin{equation}
S_{q}=\left\langle \ln_{q} \left( \frac{1}{p}\right) \right\rangle _{%
\text{lin}}
\end{equation}

\noindent where $\ln_{q}(x)$ is $q$-logarithm given by

\begin{equation}
\ln _{q}(x)=\frac{x^{1-q}-1}{1-q}.
\end{equation}

\noindent In other words, Tsallis entropy preserves the linear
averaging procedure, but generalizes the definition of the
information gain $\ln (1/p)$ to the $q$-information gain
defined as $\ln _{q}(1/p)$. Then, it is evident that,
according to GEMP, the functional to be maximized must be of the
form

\begin{equation}
\Phi _{q}=\left\langle \ln_{q}  \frac{1}{p} \right\rangle _{%
\text{lin}}-\alpha \langle .\rangle _{\text{lin}}-\beta \langle
..\rangle _{\text{lin}}
\end{equation}

\noindent where $\alpha$ and $\beta$ are as usual Lagrange
multipliers associated with the linear average of 1 and the
microstate energy $\varepsilon$, respectively. Although we know
that the constraints must be written as linear averaged quantities
in accordance with the fundamental structure of the Tsallis entropy,
\textit{the form of the probability distribution to be used in the
functional is not trivial due to the generalization of the ordinary
definition of the information gain}. This is denoted by the brackets
with single and double dots. The number of dots in the brackets are
different, since the same probability distribution may not be used
in the linear averaging procedure.

Our goal is now to determine the form of the function $f(x)$ to be
used in the linear averaging procedure (since Tsallis entropy
preserves the linear averaging procedure as it is), in Eq. (19). 
One might here consider three choices: the first choice is to 
employ ordinary definition of the probability function $f(x)$, 
which simply corresponds to ordinary Fourier transform implying 
all finite moments to be calculated through 
$\langle x^{n}\rangle _{\text{lin}}=\int dxf(x)x^{n}$, $(n=0,1,...)$. 
This is called first choice of constraints in the literature [10,18]. 
However, the second and third choices of constraints are related to 
$q$-Fourier transform [32] instead of the ordinary one, since a function 
$f(x)$ can alternatively be determined by its $q$-moments as a result 
of rewriting the definition of the information gain through $q$-logarithm. 
Within this scheme, the third choice can easily be obtained as follows: 
the $q$-Fourier transform [32] of probability density $f(x)$ is given by

\begin{equation}
F_{q}[f](\xi )=\int\limits_{-\infty }^{+\infty }dxe_{q}(i\xi
x[f(x)]^{q-1})f(x),q\geq 1
\end{equation}

\noindent where the $q$-exponential i.e., $\exp_{q}(x)$ is 

\begin{equation}
\exp _{q}(x)=[1+(1-q)x]_{+}^{\frac{1}{1-q}} . 
\end{equation}

\noindent Then, it is possible to show that

\begin{equation}
\frac{1}{\upsilon _{q_{n}}}\left[ \frac{d^{(n)}F_{q}[f](\xi )}{d\xi ^{n}}%
\right] _{\xi =0}=i^{n}\left\{
\prod\limits_{m=0}^{n-1}[1+m(q-1)]\right\} \langle x^{n}\rangle
_{q_{n}},n=1,2,...
\end{equation}

\noindent where $q_{n}=1+n(q-1)$ and $q_{n}$-mean moments $\langle
x^{n}\rangle _{q_{n}}$ are given by

\begin{equation}
\langle x^{n}\rangle _{q_{n}}=\frac{\int dxx^{n}f^{q_{n}}}{\int
dxf^{q_{n}}}
\end{equation}

\noindent with

\begin{equation}
\upsilon _{q_{n}}=\int\limits_{-\infty }^{+\infty }dx[f(x)]^{q_{n}}.
\end{equation}

\noindent This result obtained by Tsallis \textit{et al.} [32] is
very important from our point of view, since, as emphasized by
Tsallis \textit{et al.} too, it shows that one must use the
following generalized escort distributions

\begin{equation}
f_{q_{n}}(x)=\frac{[f(x)]^{q_{n}}}{\int
dx[f(x)]^{q_{n}}}, \quad n=0,1,2,...
\end{equation}

\noindent whenever one needs to calculate an averaged quantity in
$q$-space. The results of Tsallis \textit{et al.} [32] can be summarized 
as follows: the probability density that must be used in Eq. (19) is the 
generalized escort distributions given by Eq. (25). 
If we want to obtain the linear average of a constant, we
substitute $n=0$ in Eq. (25). For any first moment, we substitute
$n=1$ and so on. Therefore, we write, for the linear average of $1$
in the $q$-space, as

\begin{equation}
\langle 1\rangle _{\text{lin}}=\int dxf_{q_{0}}(x)\times 1=\frac{\int dx[f(x)]}{%
\int dx[f(x)]}=1.
\end{equation}

\noindent The inspection of the above equation shows us that the
normalization has to be carried out in terms of the ordinary
probability density function $f(x)$. Next, we consider the linear
average of the microstate energy $\varepsilon(x)$ in $q$-space

\begin{equation}
\langle \varepsilon\rangle _{\text{lin}}=\int
dxf_{q_{1}}(x)\times \varepsilon =\frac{\int dx\varepsilon \lbrack
f(x)]^{q_{1}}}{\int dx[f(x)]^{q_{1}}}=\frac{\int dx\varepsilon
\lbrack f(x)]^{q}}{\int dx[f(x)]^{q}}.
\end{equation}

\noindent At this point, it should be emphasized that the linear 
average above is taken in terms of the generalized distribution 
given by Eq. (25) as a result of employing $q$-moments. 
The maximization of the functional in Eq. (19), subject to
constraints Eqs. (26) and (27), yields the well-known
stationary distribution

\begin{equation}
p=\frac{[1-(1-q)\beta (\varepsilon-U_{q})/\int
dxp(x)^{q}]^{1/(1-q)}}{Z_{q}}
\end{equation}

\noindent where $U_{q}=\langle \varepsilon\rangle
_{\text{lin}}$, and the partition function $Z_{q}$ is given by

\begin{equation}
Z_{q}=\int dx\left[1-(1-q)\beta \frac{(\varepsilon-U_{q})}{\int dxp(x)^{q}}%
\right]^{1/(1-q)}.
\end{equation}

\noindent  Eq. (28) is the well-known probability distribution associated with 
the third choice of constraints [18].
Moreover, the probability distribution associated with the second choice of constraints 
can be obtained from the formalism above solely by using unnormalized $q$-moments.

Summing up, the algebra underlying the first choice is ordinary Fourier transform 
with $q'$, whereas the second and third choices require the use of $q$-Fourier 
transform with $q$, satisfying $q'=2-q$ [31, 33]. 
Therefore, all choices used so far in nonextensive statistical mechanics [18] 
emerge naturally within GEMP.

\section{Conclusions}
The emergence of the generalized entropy measures rendered the
generalization of the maximization procedure necessary. These
generalized entropy measures generally stem from the interplay of
the generalization of the information gain and/or averaging
procedure. For instance, the R\'{e}nyi entropy is obtained as a
generalization of BG entropy through the generalized averaging
procedure i.e., the exponential average, whereas Tsallis entropy is
a generalization through a novel definition of information gain.
Despite all these generalizations of the entropy measures in terms
of information gain and/or averaging procedures, the maximization
procedure has in general been applied without taking these changes
into account properly. We therefore proposed a generalized entropy
maximization procedure, which takes into account the averaging
procedure and information gain underlying the generalized entropies.
This novel procedure was then applied to the R\'{e}nyi and Tsallis
entropies.

Since the R\'{e}nyi entropy is a generalization of BG entropy in
terms of exponential averages, the generalized maximization
procedure requires the consistent use of the exponentially averaged
constraints. This novel procedure applied to the R\'{e}nyi entropy
then yields a stationary distribution, which asymptotically decays
as an exponential for $q\in \lbrack 0,1]$, instead of inverse power
law distributions obtained from the ordinary maximization procedure.
It should be noted that this result is in complete agreement with
the one obtained from the method of multinomial coefficients [30].
Moreover, the inspection of the stationary distribution obtained
from the maximization of the R\'{e}nyi entropy shows that it
asymptotically attains a constant value, independent of the
microstate energy $\varepsilon _{i}$, for $q>1$. Therefore, the
R\'{e}nyi measure cannot be used as a thermodynamical entropy in the
region $q>1$. This interval was generally excluded by recourse to
the fact that the R\'{e}nyi measure is neither concave nor convex in
the aforementioned interval. The consistent maximization of the
R\'{e}nyi measure in this work simply shows the underlying reason
why this measure cannot be used for the region $q>1$ at the level of
the stationary distribution. The stationary solution obtained
through ordinary maximization of the R\'{e}nyi measure, on the other
hand, seems to be valid for all $q$ values.

The nonadditive Tsallis entropy preserves the linear averaging
procedure in its definition, but deforms the definition of the
information gain. As a result, there exist two possibilities: 
the first one is to carry out the averaged constraints in terms of 
moments determined by the ordinary Fourier transform, whereas the 
second possibility is to make use of $q$-moments, stemming from the 
$q$-Fourier transform, in the averaging procedure. 
The former corresponds to the probability distribution associated 
with the first choice of constraints. On the other hand, the 
probability distribution associated with the third choice of constraints 
is obtained from the latter. 
Finally, the use of unnormalized $q$-moments results in the 
probability distribution related to the second choice of contraints. 
In other words, in our unifying scheme, all choices of constraints 
naturally emerge depending on the adoption of either ordinary or 
$q$-deformed Fourier transforms. 
The resulting stationary distributions, in all cases, are genuine inverse 
power laws.

\vspace*{-0.5cm}
\section*{Acknowledgments}
\vspace*{-0.5cm}
We are grateful to C. Tsallis for bringing Ref. [32] to our
attention. This work has been supported by TUBITAK (Turkish Agency)
under the Research Project number 108T013.


\end{document}